\documentclass[showpacs,amsmath,amssymb,floatfix,prl,twocolumn]{revtex4}

\usepackage[pdftex]{graphicx}
\usepackage{amssymb,amsfonts,amsmath}
\usepackage{graphicx,amsmath}
\usepackage{float}

\begin{document}

\title{Probing the localization length of photo-generated charges in organic materials.}
\author{A.D. Chepelianskii, J. Wang and R.H. Friend \\
 Cavendish Laboratory, University of Cambridge, J J Thomson Avenue, Cambridge CB3 OHE, UK
}


\begin{abstract}
We report a new experimental method to measure the localization length of photo-generated carriers in an organic donor-acceptor photovoltaic blend
by comparing their dielectric and electron spin-resonance susceptibilities which are simultaneously measured 
by monitoring the resonance frequency of a superconducting resonator.
We show that at cryogenic temperatures excitons are dissociated into long lived states, but that these are confined within a separation of around $4\;{\rm nm}$.
We determine the Debye  and recombination times, showing 
the coexistence of a fast electrical response corresponding to delocalized motion, with glass-like 
recombination kinetics.
\end{abstract}

\maketitle

Organic photovoltaic cells, OPVs, rely on heterojunction between donor and acceptor materials 
to efficiently separate excitons into free charges \cite{YuHeeger,Durrant}. 
Understanding the charge generation mechanisms,
and controlling the nano-scale architecture of these materials is crucial 
to the improvement of their power conversion efficiencies. Research in this direction has stimulated 
creative experiments combining material optimization, electrical transport and optical spectroscopy \cite{Bradley,Manca,Brabec,Vardeny,Marsh,Artem}.
At room temperature,
thermal activation allows carriers to hop through amorphous regions 
or domain boundaries. In contrast at low temperatures hopping is strongly suppressed and, as we show here,
carriers become confined within small nm length scale domains. 
In this Letter, we investigate the transport properties of photo-excited carriers 
in this regime. Our results provide an insight on the typical length scales over which 
carriers are delocalized in absence of thermally activated transport. 

\begin{figure}
\centerline{\includegraphics[clip=true,width=8cm]{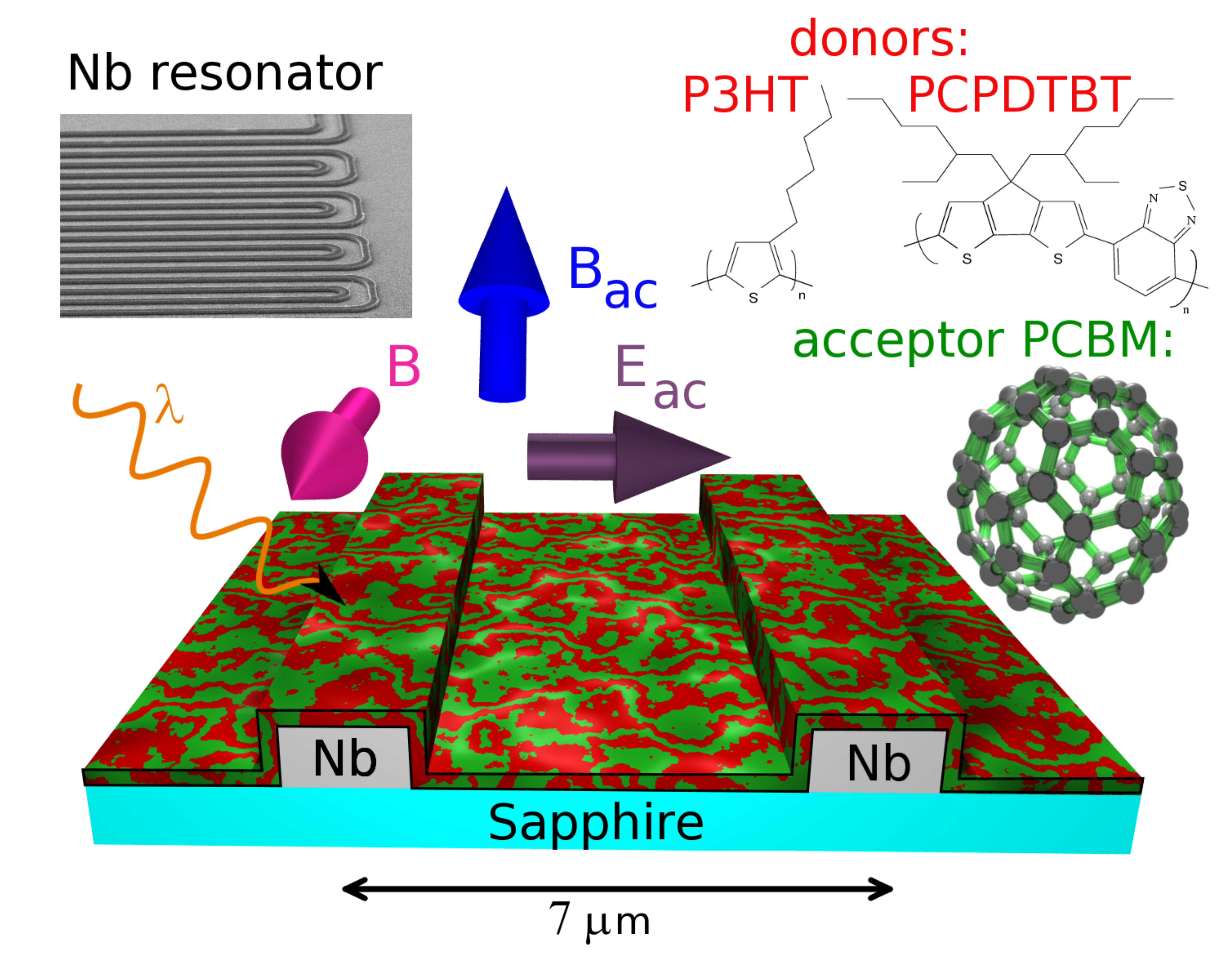}}
\caption{Schematic of the experiment: $Nb$ meanders are covered by a thin PV layer
which is coupled to the AC electric $E_{ac}$ and magnetic $B_{ac}$ fields produced by the resonator.
The sample can be optically excited at wavelength $\lambda$ and a static uniform magnetic field $B$ can also be applied.
Top left panel shows a scanning electron microscope image of the resonator and the chemical structures 
of the used molecules are summarized on the right (a $C_{60}$ molecule represents the fullerene derivatives).
}
\label{FigSetup}
\end{figure}

Due to the confined nature of carriers at cryogenic temperatures, it is necessary to use a contact-less technique
to access their properties, as achieved in a microwave domain experiment \cite{Warman,Holt}. 
Since the optical absorption length in organic materials is usually in the 100\;{\rm nm} range,
standard microwave cavities are not ideally suited for investigation of photo-transport properties, 
as photo-excitations can be generated only in a small fraction of the cavity volume.
However, at low temperatures, superconductive strip-line resonators confine 
the electromagnetic field on a few microns scale, while achieving very high quality factors.
These considerations led us to the experimental setup that is summarized in Fig.~\ref{FigSetup}.
The resonator consists of $2\;{\rm \mu m}$ wide Nb meanders lying on a transparent Sapphire substrate \cite{Bouchiat1,Bouchiat2,Chiodi,Bastien}. 
It operates at a fundamental frequency of $f_0 \simeq 365\;{\rm MHz}$. 
A typical photovoltaic blend is deposited on the surface by spin coating.
The superconducting meanders create AC electric/magnetic fields whose direction is mainly
parallel/perpendicular to the organic layer. In addition a static magnetic field can be applied 
along the direction of the meanders. The entire system is placed in an optically accessible cryostat
allowing optical excitation with a $\lambda = 532\;{\rm nm}$ laser or a monochromator
(using light from a broadband Xenon source). In most experiments, the sample was immersed in Helium-II  
providing highly efficient thermalization to the bath temperature $T = 2\;{\rm K}$.
The materials investigated in this study are donor-acceptor blends P3HT:PCBM and PCPDTBT:PC(70)BM \cite{chem} 
with respective polymer/PCBM weight ratios of 1:1 and 1:2 annealed in nitrogen atmosphere at 
$130^{{\rm o}}\;{\rm C}$ \cite{Brabec,Heeger2}.

\begin{figure}
\centerline{\includegraphics[clip=true,width=8cm]{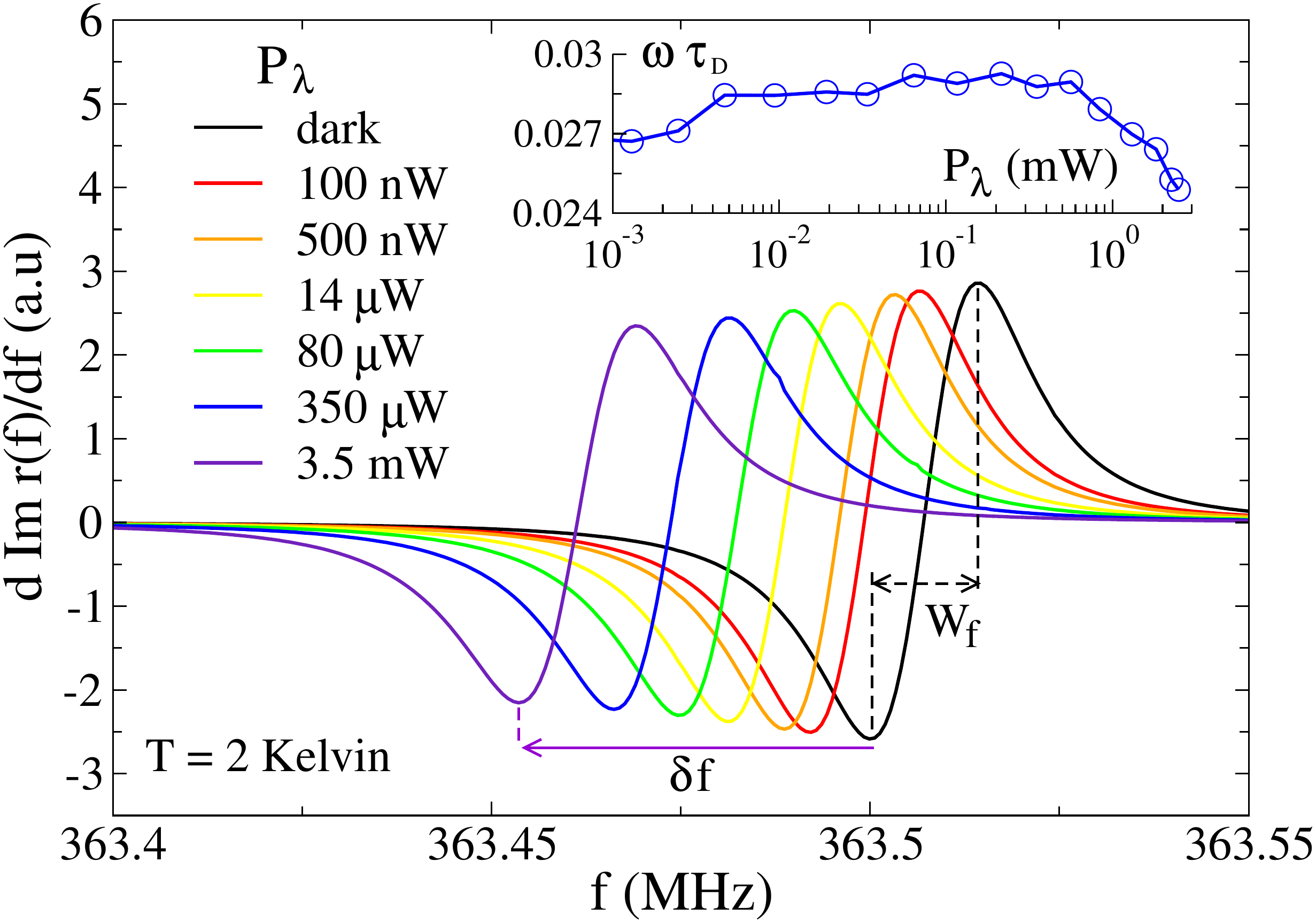}}
\caption{Displacement of the cavity resonance towards lower frequencies under illumination for a PCPDTBT:PC(70)BM layer.
The change in the lineshape can be described by the displacement of the resonant frequency $\delta f$ and a weaker 
variation of the linewidth $W_f$. The inset shows the pump power dependence of the product $\omega \tau_D$ which is deduced from the dimensionless ratio $\delta W_f/\delta f$. Typical microwave power was $1\;{\rm nW}$.
}
\label{FigDeltaF}
\end{figure}

In a first set of experiments, we measured the microwave reflection signal $r(f)$ from the 
resonator as a function of the probe frequency $f$. The reflection coefficient was measured 
with a mixer that multiplied the reflected wave from the resonator with a reference signal with an adjustable phase.
In order to increase sensitivity, a frequency modulation (FM) was applied to the microwave probe (amplitude of $5\;{\rm kHz}$, and modulation frequency $713\;{\rm Hz}$). The output voltage from the mixer was measured at the FM modulation frequency with a lock-in amplifier giving a signal proportional to $d\;{\rm Im}\;r(f)/df$. This quantity is displayed 
on Fig.~\ref{FigDeltaF} as a function of the probe frequency $f$, in the dark and at various illumination powers.
This frequency dependence gives the position of the resonance frequency $f_0$ and the quality factor 
of the resonator which depends on the peak to peak distance $W_f$.
All the line-shapes are well described by the derivative of a Lorentzian. 
Under illumination, the resonant frequency decreases by an amount $\delta f$ 
accompanied by a less noticeable broadening of the resonance by an amount: $\delta W_f$.
The control experiments without the polymer 
layer exhibited frequency shifts that were several orders of magnitude smaller, 
the dependence of the frequency shift $\delta f$ on the pump wavelength $\lambda$ matched the optical absorption 
spectrum of the PV materials, and illumination of pristine materials (for e.g. P3HT only) 
lead to much smaller effects as compared to the blends.

Since the frequency shift dominates over the broadening $\delta f \gg \delta W_f$ 
the photo-generated carriers are probably localized. 
Indeed such carriers contribute mainly to the dielectric signal resulting in an increase 
in the capacitance which lowers the resonant frequency \cite{BlueRay}. 
This contrasts with the reported room temperature 
behavior where the broadening effects related to photo-conductivity are dominant \cite{Warman,Holt}.
Information on the time scale for the motion of the confined carriers can be 
obtained from the ratio between $\delta W_f$ and $\delta f$.
For confined carriers whose polarizability $\alpha_e(\omega = 2 \pi f)$ 
is described by a Debye response $\alpha_e(\omega) = \alpha_{e}/(1 + i \omega \tau_D)$  (we confirmed this 
by measuring $\delta f$ at several harmonics of the resonator)
we find that: $\delta W_f/ \delta f = 2 \omega \tau_D /\sqrt{3}$ where $\tau_D$, the Debye time, 
is the characteristic response time of the carriers, it can be determined directly from Fig.~\ref{FigDeltaF}.
The inset shows that $\tau_D$ depends only weakly on the optical excitation power. 
We find $\tau_D \simeq 18\;{\rm ps}$ for P3HT:PCBM and $\tau_D \simeq 12\;{\rm ps}$ for PCPDTBT:PC(70)BM.
These time-scales are substantially longer than the vibrational/relaxational time scales expected 
for excited states on a single molecule that are typically in the $100\;{\rm fs}$ range.
Thus the carriers are probably delocalized on meso-scale molecular clusters,
however the effective localization length cannot be determined directly from $\tau_D$.

This length-scale can be obtained from the dielectric polarizability $\alpha_e$ per carrier,
which can be expressed as $\alpha_e = \epsilon_0 a_l^3$ where the quantity $a_l^3$ is the polarization volume and 
$\epsilon_0$ is the vacuum permittivity. We have defined the polarizability without introducing the dark value of the PV film 
dielectric constant $\epsilon_r$ since it does not influence the electric field inside the layer in our geometry.
Note also that $\epsilon_r$ comes from the polarizability of fully occupied molecular orbitals and vibrational 
states, whereas here we investigate the contribution to the polarizability from charged excitations.
We term the associated length-scale $a_l$  the localization length, 
sometimes also called the electrostatic radius.
For example $a_l$ would be approximately the Bohr-radius for a hydrogen-like atom, or the molecular radius for the 
polarizability of fullerenes \cite{Gueorguiev}. Using the electromagnetic theory of hollow resonators \cite{Landau} we can find the relative frequency 
shift $\delta f_{el}/f$ (the suffix in $\delta f_{el}$ highlights its dielectric origin) created by $N_S$ carriers distributed on the surface of the resonators 
\begin{align}
\frac{\delta f_{el}}{f} = -\frac{N_s a_l^3}{S_R \lambda_E} 
\label{eq:deltaf}
\end{align}
where $S_R$ is the surface area of the resonator and $\lambda_E$ is the effective confinement length for the AC electric field. 
In our case, we find $S_R = 2.8\;{\rm mm^2}$ and $\lambda_E \simeq 19.8\;{\rm \mu m}$ as 
obtained by finite-element simulations of the field profiles;
as a consistency check we verified that our simulations gave the accurate value for the fundamental resonance frequency $f_0$.
Note also that $a_l^3$ corresponds to the average polarization volume between electrons and holes. 
Even if the left hand side of Eq.~(\ref{eq:deltaf}) is known from the data on Fig.~\ref{FigDeltaF}, 
this equation cannot be used to determine $a_l^3$ directly without knowledge of the photo-induced carrier population $N_S$.
Since the charges of the photo-induced electrons and holes compensate each other, $N_S$ cannot be obtained 
from a charge measurement as in a field-effect transistor. However each carrier has a spin $1/2$ and 
so it is possible to count the total number of spins $N_S$ by measuring the light induced electron-spin resonance (LESR) signal
\cite{Dyakonov,Tanaka}.

\begin{figure}
  \centerline{\includegraphics[clip=true,width=8cm]{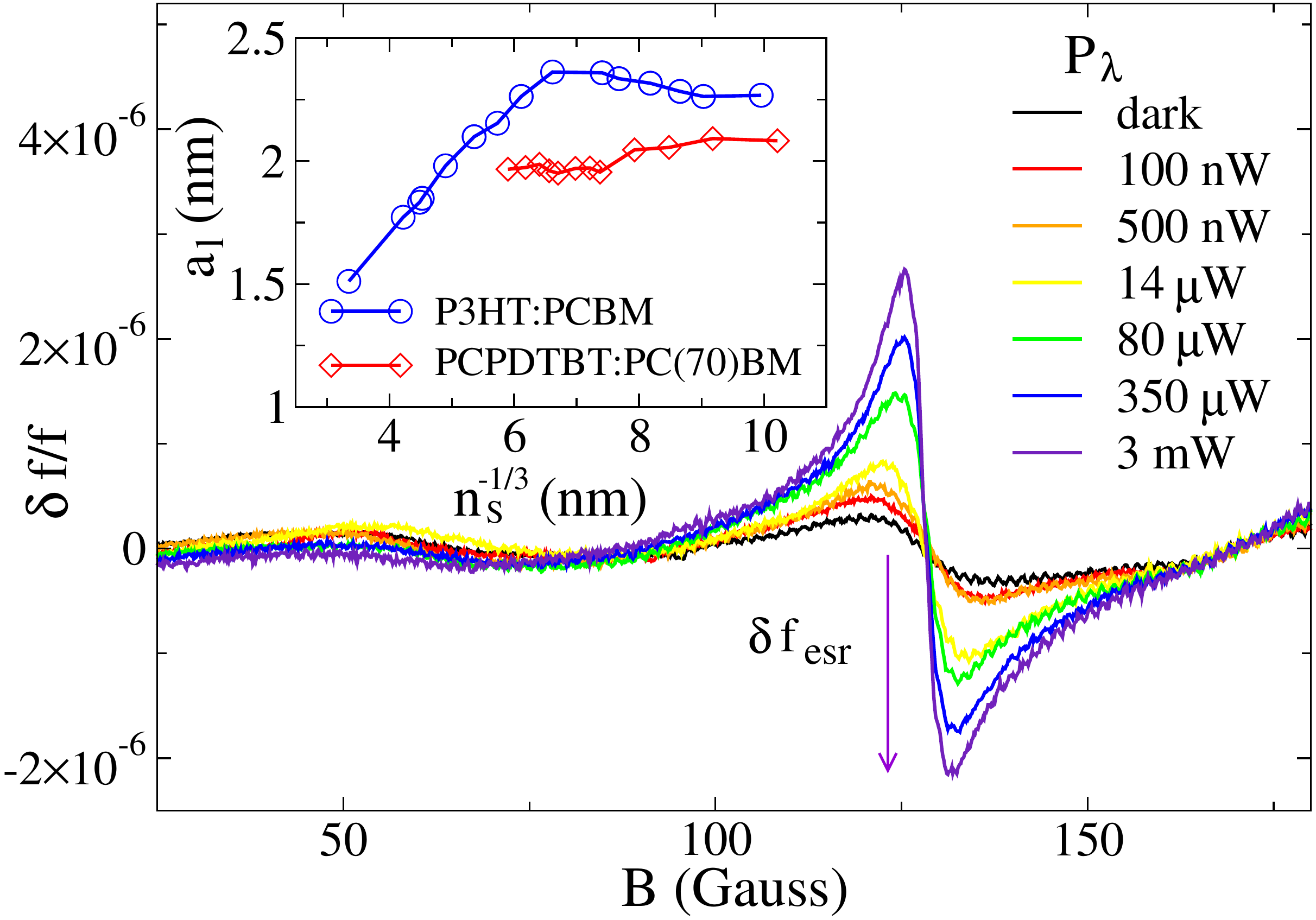}}
\caption{Magnetic field dependence of the resonant frequency shift $\delta f/f$ at several illumination powers for a PCPDTBT:PC(70)BM layer.
The LESR signal is centered at $B \simeq 130\;{\rm Gauss}$ with an increasing amplitude as a function of $P_\lambda$. 
Similar data was obtained for P3HT:PCBM. The inset shows the deduced localization length $a_l$
as a function of the separation between carriers $n_S^{-1/3}$ (where $n_S$ is defined as $n_S = N_S/(S_R W)$).
}
\label{FigDeltaESR}
\end{figure}

The LESR is detected by monitoring the resonant frequency of the cavity as a function 
of a static magnetic field $B$ applied as indicated on Fig.~\ref{FigSetup}.
Since in our resonator both the electric and magnetic AC fields ($E_{ac}$/$B_{ac}$) interact
with the PV layer, the AC magnetic field 
can drive the electron-spin resonance (ESR) producing a displacement
of the resonance frequency when it matches the ESR frequency:
 $\omega_B \simeq g \mu_B B$, where $g \simeq 2$ is the $g$ factor and $\mu_B$ is the Bohr-magnetron.
The dependence of the relative frequency shift of the cavity $\delta f/f$ on the magnetic field $B$ 
is displayed in Fig.~\ref{FigDeltaESR}. It shows a clear LESR resonance
appearing as expected at $B = 2 \pi f/(g \mu_B)\simeq 130\;{\rm Gauss}$ for a resonant frequency $f \simeq 363MHz$.
As expected LESR amplitude increases with the optical excitation power.  
The absence of zero field splitting and of a half field transition 
at $B/2 \simeq 65\;{\rm Gauss}$ in the LESR signal 
confirms our identification of the excited states as separated electron-hole pairs.
For clarity, a parabolic dependence of  $\delta f$ on the magnetic field $B$ (independent on the illumination power) 
has been subtracted from the data \cite{Bouchiat1}.

The number of photoexcited spins $N_S$ can be estimated from LESR 
using the known magnetic susceptibility $\chi_m$ of a paramagnetic spin $1/2$ and the effective
confinement length for the AC magnetic field $\lambda_B \simeq 5.6\;{\rm \mu m}$ :
\begin{align}
\frac{\delta f_{esr}}{f} = \frac{N_S \chi_m}{S_R \lambda_B} = \frac{N_S}{S_R \lambda_B} \frac{\mu_0 g^2 \mu_B^2}{16 k_B T}   \frac{\omega_B (\omega_B - \omega) \tau_2^2}{1 + (\omega_B - \omega)^2 \tau_2^2} 
\label{Eq:ESR}
\end{align}
where $\tau_2$ is the spin dephasing time \cite{ESRbook}.
Note that we use the real part of the magnetic susceptibility since we  
measured the shift of the resonant frequency and not the cavity losses. We found that Eq.~(\ref{Eq:ESR})
provided a good description of our ESR data with $\tau_2$ ranging from $10$ to $20\;{\rm ns}$ 
which was used as a fit parameter. These values are consistent with the linewidths reported 
in optically detected magnetic resonance experiments \cite{Shinar}. The theoretical spin sensitivity was confirmed 
by depositing a reference paramagnetic on the resonator \cite{DPPH}.
Combining Eqs.~(\ref{eq:deltaf},\ref{Eq:ESR})
we find the ratio between dielectric and magnetic signals $\delta f_{esr}/\delta f_{el} = \chi_m \lambda_E /(a_l^3 \lambda_B)$.
In the limit of a thin Nb film the coupling ratio between electric and magnetic modes is simply 
$\lambda_E/\lambda_B = (1 + \epsilon_S)/2$ where $\epsilon_S = 10$ is the dielectric constant of Sapphire,
however the finite $1\;{\rm \mu m}$ thickness of the Nb film leads to a lower numerical value $\lambda_E/\lambda_B \simeq 3.55$. 

The above results 
allowed us to determine the localization length as a function of the light induced spin density $n_S = N_S/(S_R W)$
where $W \simeq 130\;{\rm nm}$ is the PV film thickness and $N_S$ was deduced from ESR.
The spin density $n_S$ is also the sum of the photo-generated electron and hole densities $n_S = n_e + n_h$.
Fig.~\ref{FigDeltaESR} (inset) shows the variation of 
$a_l$ as of a function of the mean photo-generated carrier separation $n_S^{-1/3}$,
revealing a plateau at large values of $n_S^{-1/3}$. The plateau is centered around
$a_l \simeq 2.3\;{\rm nm}$ for P3HT:PCBM and $a_l = 2.1\;{\rm nm}$ for PCPDTBT:PC(70)BM.
In the case of P3HT:PCBM the mean carrier separation $n_S^{-1/3}$ and the 
the localization length $a_l$ become comparable at the highest illumination powers.
A drop of the polarization volume is observed this regime, as expected due to the increasing strength of the Coulomb 
interactions at high densities \cite{Screening}. The crossover occurs when $n_S^{-1/3} \sim a_l$,
providing an additional consistency check since the two quantities $n_S^{-1/3}$ and $a_l$ were determined independently.
We note that the same dependence was measured by exciting the P3HT:PCBM sample with a $407\;{\rm nm}$ laser.

\begin{figure}
\centerline{\includegraphics[clip=true,width=8cm]{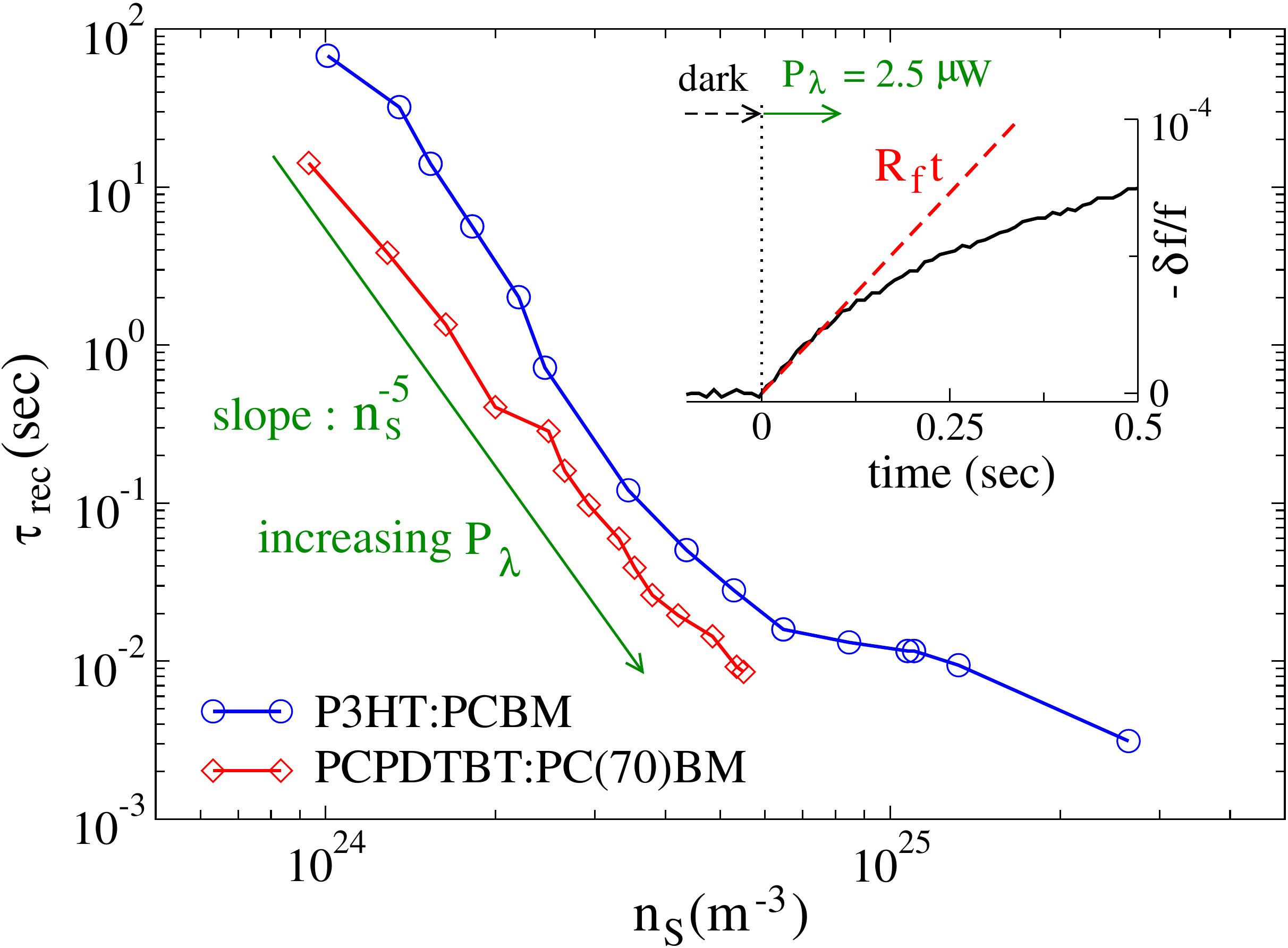}}
\caption{Mean recombination time $\tau_{rec}$ as a function of the total carrier 
density $n_S$ as determined from Eq.~(\ref{eq:TauRec}). The inset shows a typical trace where the frequency shift rate $R_f$ is measured, 
when an illumination of known power $P_{\lambda} = 1.5\;{\rm \mu W}$ is switched on. 
}
\label{FigTauRec}
\end{figure}

It is interesting to compare the values of $a_l$ which we report here 
and our knowledge on the blend morphology. There is currently a growing evidence in support of a structure 
consisting of small quasi-crystalline PCBM clusters of a few nanometer radius 
embedded in a mixture of amorphous and crystalline polymer \cite{Wei,Bao,Liao,Marletta,Sven}. 
Thus it seems likely that the photo-polarizability effect
originates from delocalized carriers inside the fullerene nanocrystals, consistent with the similar values found in
P3HT:PCBM and PCPDTBT:PC(70)BM. In this scenario $a_l$ would be determined 
by the size of the fullerene nanocrystals, suggesting the exciting perspective 
of combining our novel technique with  material science methods to control in-situ the degree of phase separation 
or the size of the fullerene clusters. 

Knowing the average polarizability per photo-generated carrier, it was possible
to probe the carrier generation efficiency and the recombination time-scales at low temperatures.
The following experiment was used to determine the exciton separation quantum efficiency (ESQE) 
in our samples: the resonator was kept in the dark for a long enough time to allow relaxation of 
most of the carriers, then at time $t = 0$ an irradiation of known intensity $P_\lambda$ was switched on
leading to a transient decrease of the resonance frequency at a rate $d (\delta f/f) /d t = - R_f$.
A typical experimental trace is shown in the inset on Fig.~\ref{FigTauRec}, note that $R_f$ was measured 
immediately after the illumination started to avoid contributions from later time recombination effects 
which lead to the stabilization of the resonance frequency.
The rate $R_f(P_\lambda)$ was proportional to the incident power $P_\lambda$, for example 
for P3HT:PCBM the proportionality held as $P_\lambda$ was varied by three orders of magnitude 
 in the range $1.5\;{\rm nW} - 1.5\;{\rm \mu W}$. Thus we can interpret $R_f(P_\lambda)$ as the product between the rate at 
which electron-hole pairs are generated and the relative drop of the resonant frequency per carrier which we found to be $a_l^3/(S_R \lambda_E)$ based on electromagnetic modeling. This leads to the following relation: 
\begin{align}
-\frac{d (\delta f/f)}{d t} = R_f(P_\lambda) = \frac{a_l^3}{S_R \lambda_E} (2 \times {\rm ESQE}) \frac{P_\lambda}{\hbar \omega_\lambda}
\label{eq:ESQE}
\end{align}
where we defined ESQE as the probability of forming a charge separated electron-hole pair per incident photon.
The latter are generated at a rate $P_\lambda/(\hbar \omega_\lambda)$ where $\hbar \omega_\lambda$ is the photon-energy.
All the quantities appearing in Eq.~(\ref{eq:ESQE}) are known except for ESQE,
thus we can use it to find ${\rm ESQE} = 12.8\%$ for P3HT:PCBM and ${\rm ESQE} = 5.6\%$ PCPDTBT:PC(70)BM.
These values are substantially lower than those reported at room temperature,
however the probability of an exciton to diffuse to a donor-acceptor interface is smaller at low temperatures
and the germinate recombination of charge-transfer states is also enhanced.
The possibility to study charge separation at low temperature under low intensity irradiation opens up the possibility to discriminate between the ``hot'' and ``cold'' charge generation mechanisms which are still debated \cite{Troisi}.

We have also investigated the dependence of the recombination life-time $\tau_{rec}$ on the carrier density.
This quantity was determined from the steady state balance between the generation term $R_f(P_\lambda)$  
and the decrease in polarizability due to recombination of the carriers at a rate $N_S / \tau_{rec} = (n_S W S_R)/\tau_{rec}$ :
\begin{align}
R_f(P_\lambda) = \frac{a_l^3 W}{\lambda_E} \frac{n_S}{\tau_{rec}} = -\frac{1}{\tau_{rec}} \frac{\delta f_{el}}{f}
\label{eq:TauRec}
\end{align}
the obtained results are displayed in Fig.~\ref{FigTauRec}.
We note that the lowest values of $\tau_{rec}$ are in the ${\rm ms}$ range, several orders of magnitude larger than 
the Debye time $\tau_D \sim 10\;{\rm ps}$.
This is consistent with our interpretation that electron and holes are separated by a barrier
with an exponentially slow hopping rate preventing recombination. Interestingly the lifetimes in P3HT:PCBM 
are consistently larger by almost an order of magnitude as compared to PCPDTBT:PC(70)BM. 
A steep dependence on density is observed $\tau_{rec} \propto n_S^{-5}$ in both materials,
in a Langevin model we would expect $\tau_{rec} \propto n_S^{-1}$.
It corresponds to a high order recombination kinetic, suggesting that the effect of interaction between carriers 
are more pronounced at low temperatures and require a treatment beyond the bi-molecular interactions 
included in the Langevin model.
The fast divergence of the recombination times at low carrier density
suggests an analogy with the ``electron glass'' phase \cite{Grenet},
to confirm this analogy, we monitored the relaxation of the resonant frequency when illumination was stopped 
and finding a $\log t$ time tail characteristic of a glassy system. Other effects can also lead to 
this unusual power law dependence, for example the percolating nature of the conducting paths in disordered conductors
\cite{Percolation} or tail states inside the gap \cite{TailStates}. An interesting consequence of the existence 
of these very long lived carriers is that charge extraction remains possible even at cryogenic temperatures.
We have measured the external quantum efficiency in OPV devices at $T = 10\;{\rm K}$ finding values 
in the $0.5\%$ range \cite{Wang}, 
which indicates that the long lived photo-generated charges reported here can be converted into photo-current. 

In summary, we have introduced a new technique to probe the localization length of photo-excited carriers.
This allowed us to measure the localization length in two reference organic PV materials leading to a value of around $2\;{\rm nm}$ 
compatible with the size of fullerene nano-crystalline domains. 
We have also shown that charge separation
is possible even at cryogenic temperatures and that it is not necessarily thermally activated. 
The generated carriers are extremely long lived, exhibiting an unusual dependence of the recombination time 
on the carrier density, thus they are likely to be separated by at least twice the localization length or $4\;{\rm nm}$.

We thank H. Bouchiat for fruitful discussions and A. Barnett for 
assistance in the experiment preparation. We thank the EPSRC for support and A.C. acknowledges the E. Oppenheimer Foundation and St Catharine's College, Cambridge.


\begin{thebibliography}{99}
\bibitem{chem} PCBM is (3'H-Cyclopropa[8,25][5,6]fullerene-C70-D5h(6)-3'-butanoicacid, 3'-phenylmethyl ester), P3HT is poly(3-hexyl thiophene) and PCPDTBT is (poly[2,6-(4,4-bis-(2-ethylhexyl)-4H-cyclopenta[2,1-b;3,4-b']-dithiophene)-alt-4,7-(2,1,3-benzothiadiazole)])
\bibitem{DPPH} A known quantity of 2,2-diphenyl-1-picrylhydrazyl (DPPH), approximately $75\;{\rm ng}$, was deposited on the resonator. The number of coupled spins was then determined from electron-spin resonance and from the spin density in DPPH with consistent results (relative error of 4\%).
\end{thebibliography}
\end{document}